# Evaluation of Banking Sector's Development in Bangladesh in light of Financial Reform

Nusrat Jahan[1*]    K.M. Golam Muhiuddin[2]
1.Assistant Professor, School of Business, Chittagong Independent University, Bangladesh
2.Professor, Ex-Dean, AIS Department, Faculty of Business Administration, Chittagong University
* E-mail of the corresponding author: mrs_jahan@ciu.edu.bd

**Abstract**
Historically, the performance of the banking sector has been weak, characterized by weak asset quality, inadequate provisioning, and negative capitalization of state-owned banks. To overcome these problems, the initial phase of banking reform (1980-1990) focused on the promotion of private ownership and denationalization of nationalized commercial banks (SCBs). During the second phase of reform, Financial Sector Reform Project (FSRP) of World Bank was launched in 1990 with the focus on gradual deregulations of the interest rate structure, providing market-oriented incentives for priority sector lending and improvement in the debt recovery environment. Moreover, a large number of private commercial banks were granted licenses during the second phase of reforms. Bangladesh Bank adopted Basel-I norms in 1996 and Basel-II during 2010. Moreover, the Central Bank Strengthening Project initiated in 2003 focused on effective regulatory and supervisory system, particularly strengthening the legal framework of banking sector. This study evaluates how successfully the banking sector of Bangladesh has evolved over the past decades in light of financial reform measures undertaken to strengthen this sector.

## 1. Introduction

The financial sector in Bangladesh comprises the money and capital markets, insurance and pensions, and microfinance. In addition to the Bangladesh Bank the central bank of Bangladesh-there are four state-owned commercial banks (SCBs), four state-owned specialized banks (DFIs) dedicated to agricultural and industrial lending, thirty domestic private commercial banks (PCBs), nine foreign commercial banks (FCBs), and thirty one non-bank financial institutions (NBFIs). Bangladesh Bank has regulatory and supervisory jurisdiction over the entire banking sub-sector as well as the NBFIs. Most of the institutions in the financial sector are characterized by a mix of public and private ownership. During the first decade of independence, financial system of Bangladesh has been suffering from deep crises. Historically, the performance of the banking sector of Bangladesh has been weak due to weak asset quality, inadequate provisioning, negative capitalization in systemically important state banks and constrained profitability. State banks were at times used to lend to politically important economic sectors and institutions as well as politically linked persons. Similarly, the private banks lacked sufficient checks and balances in the form of effective standards of corporate governance, resulting in high levels of related party lending. However, these trends have started to reverse in more recent times, with significant efforts made to restructure ailing state banks, improve standards of risk management, corporate governance and bank supervision and cope with international best practice standards. Following the global financial crisis of 2008 the role of financial sector regulators came under sharp scrutiny worldwide, however the banking sector of Bangladesh remain largely unaffected by the crisis. With the revamp of financial sector regulatory and supervisory frameworks with sharper focus on risk and systemic stability in line with post-global crisis and revisions of international best practice standards, the importance of conducting study on the performance evaluation of banking sector is interminable. Therefore, the aim of this study is to evaluate how successfully banking industry of Bangladesh has evolved over the past decades with regards to financial reform measures undertaken to strengthen this sector.

## 2. Financial Sector Reform Initiatives of Banking Sector

There were six nationalized commercial banks (NCBs) in Bangladesh until 1982 which are Sonali, Agrani, Janata, Rupali, Pubali and Uttara Bank. Banking system reforms in Bangladesh were required due to the serious weaknesses that threatened to destabilize the banking system, including weak asset quality and capitalization. The more severe weaknesses were at the state-owned commercial banks that were affected by poor credit underwriting standards, politically linked lending, weak management and to some extent, regulatory inaction. The initial phase of financial sector reform initiated in 1982, when the government denationalized two of the six NCBs (Rupali Bank and Pubali Bank) and permitted entry of local private banks. Rupali bank was denationalized in 1986 and transformed into PLC, with the government holding 51% shares. As of December 2007, government share in the paid up capital of Rupali bank was at 93.2%. In order to identify major problems in the financial system and to suggest remedial measures, Bangladesh government formed the 'National

Commission on Money, Banking and Credit' (NCMBC) in 1984. Many private commercial banks which were granted license in the early 1980s were Islami Bank Bangladesh (1983), United Commercial Bank (1983), City Bank (1983), National Bank (1983), Arab Bangladesh Bank (1985) and Al-Baraka Bank (1987). Despite the measures taken, initial round of reform was largely unsuccessful due to influence of vested Private Commercial Banks (PCBs) and State-owned Commercial Banks (SCBs) interest groups which resulted in loan default culture. Subsequently, a World Bank Mission conducted a comprehensive study of the financial sector and suggested reforms relating to fixation of interest rates on deposits and advances, classification of overdue loans, restructuring of capital of NCBs and PCBs and market orientation of banking transactions (Chowdhury, 2000, Robin, 2008 and Moral, n.d.).

Bangladesh Bank combined the observations and suggestions from both the NCMBC and World Bank, and undertook some initiatives aligned with these suggestions. In 1990, Financial Sector Adjustment Credit (FSAC) and Financial Sector Reform Program (FSRP) of World Bank was instituted with board objective of enhancing competitiveness in the banking industry, whereas the specific objective was to make SCBs commercially viable for subsequent privatization and help PCBs to increase their market share in total commercial banking. The policies of FSRP directed banks to provide loans on commercial basis, limited government control to the monetary policy only. It forced banks to have a minimum capital adequacy following Basel-I norms, systematically classify loans and to implement modern accounting systems and computerized systems. It forced the central bank to free up interest rates, revise financial laws and to increase supervision in the credit market. However, FSRP expired in 1996 and just before expiry government formed Banking Reform Committee (BRC) which recommended relevant acts and laws should be reviewed and amended to ensure legal enforcement which is must for attaining financial stability (Robin, 2008, Islam and Hamid, 2012).

The Banking Companies Act, 1991 was enacted in February 1991 and amended in 1993 and 1995, in order to make the role of Bangladesh Bank as authoritarian in dealing with licensing, monitoring, regulating and supervising the banking sectors. The Act deals mainly with the operations and permitted activities of the banking companies. Bangladesh Bank exercises the Act in order to reestablish the discipline in the sanctioning and rescheduling of loans and advances. Minimum capital adequacy guidelines based on Capital-to Liabilities approach was incorporated in the Banking Companies Act, 1991 and in the Banking Companies (amendment) Act, 1995. Following the recommendation of the Bank for International Settlement (BIS), Bangladesh Bank instructed all scheduled banks to submit their capital ratios and adopted Basel-I norms in 1990. In order to recover the defaulted loans, Artha Rin Adalat Act, 1990 was enacted in 1990 and a series of amendments were made to the Act in 1990, 1992, 1994 and 1997 to incorporate the effective rules to accommodate the changed situation. Prior to setting up of Artha Rin Adalat, there was no special law for recovery of loans. Even after so many amendments, the Act still suffers from some drawbacks. The Bankruptcy Act, 1997 has been enacted in March 1997. The Act has been enacted with a view to judge a debtor as bankrupt, make smooth realization of a bankrupt's assets, discharge the debtor after distribution of his assets, and take effective measures against defaulting borrowers (Chowdhury, 2000).

As per recommendation of FSRP, Bangladesh bank introduced a flexible market-oriented interest rate policy in 1990 abolishing the earlier system of centrally administered interest rate structure and sector specific concession refinancing facilities. Under the new policy, interest-rate bands were prescribed for different categories of loans, advances and deposits within which banks are free to fix their own rate. Interest rate spreads were closely monitored by Bangladesh banks and banks are advised to keep the spread below five percent due to concern over hampered credit allocation to priority sectors of the economy. Bangladesh Bank has established Credit Information Bureau (CIB) in August 1992 in order to provide reliable information regarding the credit worthiness of borrowers. The aim of CIB report is to avoid duplication of credit facilities, avoid credit facilities to defaulters and justify the status of the borrowers. In order to ensure early recognition of Non-Performing Loans, Bangladesh Bank has adopted the Loan Classification and Provisioning policy formulated by the Banking Control Division (BCD) in 1989 in order to attain international standard. All the commercial banks have to set up required provisions for their loans. According to this policy, a certain portion of the loans and advances sanctioned by all the banks have to be kept under the newly set up provisions. 'Offshore Banking Unit' (OBU) of the banks also has also to follow this new policy which made the OBU transactions more transparent. In order to manage the credit risk effectively, Bangladesh Bank has introduced the Credit Risk Grading System (CRGS) where banks have to grade their loans into eight categories. These grades are Superior, Good, Acceptable, Marginal or Watch list, Special Mention Account, Substandard, Doubtful and Loss according to the duration of the overdue status (Kabir, 2004). A supervisory unit both off-site and on-site was established in the Bangladesh Bank to evaluate the performance of banks, grading them using 'CAMEL' rating system, a device judging five major indicators of banks on a scale of 1 to 5 which are capital adequacy, asset quality, management, earning and liquidity. (Robin, 2008)

During May 1997, government undertook another project, namely "Commercial Bank Restructuring Project (CBRP)" also funded by the World Bank. This project was undertaken to identify urgent course of actions needed for continuing the pace and progress so far done. Second phase of reform though resulted in greater private participation and development of wide range of financial products and services but the banking sector did not generate expected result until the late 1990s. However, earlier reform measures undertaken failed to overcome some key problems such as high nonperforming loan in both SCBs and PCBs and enforcement of capital adequacy and other regulatory requirements. Therefore, in early 2000s, reform measures are shifted to risk-based regulations and supervision of banking sector. World Bank submitted some recommendations which are "effective legal system, good management, and effective central bank" as three pillars of banking and proposed to rebuild these pillars first. The World Bank urged to go for privatization only after the successful completion of financial restructuring of the SCBs. The Word Bank identified less attractive pay structure of SCB officials, excessive influence of trade unions, absence of autonomy and accountability, poor internal governance and management, over-staffing and over-branching, and weak legal infrastructure. Based on these observations, World Bank suggested some programs mainly focusing on to improve institutional capacity, restructure SCBs, ensuring transparency, formulating legal procedure related to realize the outstanding loans, compliance with international best practices etc. Based on the recommendation of the World Bank, Bangladesh Bank undertook some reform initiatives (Rahman, 2012).

Though FSRP spelled out the issues regarding regulation and supervision but it was indeed the reform measures in post 2000s that had a main focus on risk-based banking regulations and supervision. The Central Bank Strengthening Project initiated in 2003 focused on effective regulatory and supervisory system for the banking sector, particularly strengthening the legal framework, automation and human resource development and capacity building. Money Laundering Prevention Act 2002 was enacted in April 2002 and amended in 2007. Money Loan Court 2003 has been enacted to provide speedy procedures for obtaining decrees and execution regarding settlement of loan recovery disputes. (Robin, 2008) The Enterprise Growth and Bank Modernization Project was adopted in 2004 by the World Bank to help the government achieve a competitive private banking system through a staged withdrawal through divestment and corporatisation of a substantial shareholding in the three public sector banks which are Rupali, Agrani and Janata bank and the largest state bank, Sonali. Given the government's fiscal constraints, which limit its ability to recapitalize the SCBs, what was achieved was the conversion of Sonali, Janata and Agrani Bank into PLC's by mid 2007 in order to make them more autonomous and to pursue eventual privatization (Islam and Hamid, 2012). During the conversion of three SCBs into PLC, the valuation adjustment of assets and liabilities resulted in the accumulated losses. Since the conversion of these SCBs into PLCs, notable improvements are yet to be seen. The reported gross NPL to total loan ratios for the four SCBs and DFIs were still high at 11.3% and 24.6% during 2011, indicating that significant progress still needs to be made (Anandakumar, Khan and Srivastava, 2009). However, the banking sector reform post 2000s resulted in expansion of private and foreign commercial banking activities and therefore the domination of the banking system by the SCBs is declining while PCBs and FCBs have been gaining market share in recent years, reflecting increased competition in the banking industry. Since 2002, the PCBs and FCBs have consistently outperformed state-owned specialized banks (DFIs) and SCBs in terms of growth in deposits and advances. To cope with the international best practices and to make the bank's capital more risk sensitive as well as more shock resilient, Bangladesh Bank adopted Basel II Framework in 2006 in accordance with 'International Convergence of Capital Measurement and Capital Standards' released by Basel Committee on Banking Supervision (BCBS). Basel II guidelines were made as statutory compliance for all scheduled banks in Bangladesh from January, 2010. Meanwhile, Basel-III has been published by Basel Committee for Banking Super-vision (BCBS) and Bangladesh Bank is planning to adopt the same in near future (Recent Reform Initiatives, 2012).

## 3. Objective of the Study
The broad objective of this study is to evaluate the development of banking sector of Bangladesh in light of financial reform measures undertaken during the period of 1997 to 2011. The specific objectives are:
- I) Evaluating the growth of accessibility to banking services.
- II) Examining the banking sectors contribution in financial deepening.
- III) Examining the operational efficiency of the banking sector.
- IV) Investigating the stability of the banking sector.

## 4. Research Methodology
The financial sector of a country comprises a variety of financial institutions, markets, and products. The World Bank's Global Financial Development Database developed a comprehensive yet relatively simple conceptual framework to measure financial development of two major components in the financial sector, namely the financial institutions and financial markets. According to this framework a financial sector's development should be evaluated in terms of financial depth, access, efficiency, and stability (Beck, Asli and Ross 2000). These four dimensions are measured by using some proxy indicators stated in the framework. Therefore, this study evaluates the development of banking sector of Bangladesh by using the framework given in the following Table 1.

Table 1: Financial Sector Development Framework

|  | **Financial Institutions** | **Financial Markets** |
|---|---|---|
| **Depth** | <ul><li>M2 to GDP</li><li>Deposits to GDP</li><li>Gross value added of the financial sector to GDP</li><li>Private Sector Credit to GDP</li><li>Financial Institutions' asset to GDP</li></ul> | <ul><li>Stock market capitalization and outstanding domestic private debt securities to GDP</li><li>Private Debt securities to GDP</li><li>Public Debt Securities to GDP</li><li>International Debt Securities to GDP</li><li>Stock Market Capitalization to GDP</li><li>Stocks traded to GDP</li></ul> |
| **Access** | <ul><li>Accounts per thousand adults</li><li>Branches per 100,000 adults</li><li>% of people with a bank account (from user survey)</li><li>% of firms with line of credit (all firms)</li><li>% of firms with line of credit (small firms)</li></ul> | <ul><li>Percent of market capitalization outside of top 10 largest companies</li><li>Percent of value traded outside of top 10 traded companies</li><li>Government bond yields (3 month and 10 years)</li><li>Ratio of domestic to total debt securities</li><li>Ratio of private to total debt securities (domestic)</li><li>Ratio of new corporate bond issues to GDP</li></ul> |
| **Efficiency** | <ul><li>Net interest margin</li><li>Lending-deposits spread</li><li>Non-interest income to total income</li><li>Overhead costs (% of total assets)</li><li>Profitability (return on assets, return on equity)</li><li>Boone indicator (or Herfindahl Statistics)</li></ul> | <ul><li>Turnover ratio for stock market</li><li>Price synchronicity (co-movement)</li><li>Private information trading</li><li>Price impact</li><li>Liquidity/transaction costs</li><li>Quoted bid-ask spread for government bonds</li><li>Turnover of bonds (private, public) on securities exchange</li><li>Settlement efficiency</li></ul> |
| **Stability** | <ul><li>Z-score</li><li>Capital adequacy ratios</li><li>Asset quality ratios</li><li>Liquidity ratios</li><li>Others (net foreign exchange position to capital etc)</li></ul> | <ul><li>Volatility (standard deviation / average) of stock price index, sovereign bond index</li><li>Skewness of the index (stock price, sovereign bond)</li><li>Vulnerability to earnings manipulation</li><li>Price/earnings ratio</li><li>Duration</li><li>Ratio of short-term to total bonds (domestic, int'l)</li><li>Correlation with major bond returns (German, US)</li></ul> |

Source: Beck, Asli and Ross 2000

## 5. Evaluation of Banking Sector's Development in Bangladesh
*5.1 Access to Banking Service*
Access to banking services or financial inclusion refers to the ability of individuals to access appropriate financial products and services. Chart 1 reports demographic accessibility to banking services in Bangladesh.

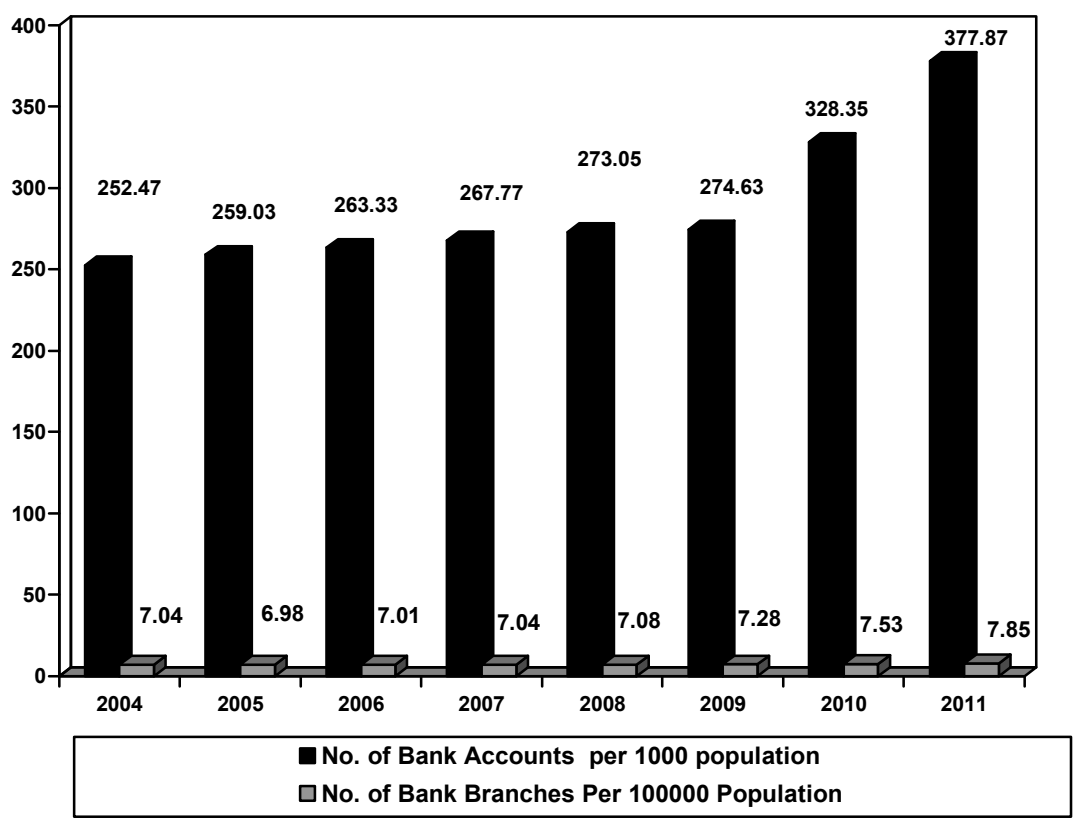

Chart 1: Demographic Penetration of Banking Services
Source: Financial Stability Report, Bangladesh Bank

Demographic penetration indicates that number of bank branches per 100,000 population increased from 7.04 to 7.85 over the period of 2004 to 2011, reporting an increment of 11.5% and number of bank accounts per 1000 population also increased by 49% over the same period. Trend in demographic penetration indicates that access to banking services has increased overtime in Bangladesh. Bangladesh government has been pursuing banks and financial institutions to follow Bangladesh Bank's financial inclusion initiatives to engage themselves in increased lending to the under-served or un-served economic sectors and population segments, including micro and SME entrepreneurs, agricultural and other rural and urban farm and non-farm productive activities for broadening access to financial services. Furthermore, in issuing new branch licenses to banks, Bangladesh Bank has been following a policy of requiring at least one in every five new branches to be in rural locations; with a view to pushing banking services physically closer to the rural population (Financial Stability Report, Bangladesh Bank).

*5.2 Financial Depth*
Financial deepening refers to the increased provision of financial services with a wider choice of services geared to all levels of society. Though it is difficult to obtain a satisfactory measure of financial deepening, it generally means an increased ratio of money supply to GDP.
*5.2.1 M3 to GDP*
This study uses ratio of some broad measures of money supply (M3) to the level of GDP to evaluate the effect of banking sector in financial deepening of a growing economy like Bangladesh. These monetization variables are used in this study to illustrate the real size of the banking sector of a growing economy in which money supply provides valuable savings and payment services. Broad money should rise at a faster rate in proportion to economic transaction if financial deepening is taking place. The ratio of M3 to GDP indicates banks ability to stimulate long-term savings relative to the level of nominal income. Bangladesh Bank begin calculating M3 from 1998, hence the Table 2 reports a rising trend in M3 to GDP ratio since 1998 and the ratio is reportedly increasing at a faster rate till 2011. Therefore, as broad money measure (M3) rising at a faster rate relative to GDP hence financial deepening is taking place in Bangladesh. This indicates that financial reform measures has geared up diversification of financial intermediation services resulting in growth of banking sector relative to the economy.

Table 2: M3 as a Percentage of GDP

| Year | M3/GDP (%) |
|---|---|
| 1998-1999 | 34.7 |
| 1999-2000 | 38.5 |
| 2000-2001 | 42.5 |
| 2001-2002 | 45.4 |
| 2002-2003 | 47.8 |
| 2003-2004 | 49.0 |
| 2004-2005 | 50.7 |
| 2005-2006 | 53.2 |
| 2006-2007 | 54.2 |
| 2007-2008 | 54.3 |
| 2008-2009 | 56.8 |
| 2009-2010 | 61.8 |
| 2010-2011 | 64.8 |

Source: Economic Research

*5.3 Efficiency*

The proponents of financial reform argue that efficient operation of financial institution is prerequisite to financial sector development. Following framework of World Bank, indicators which are evaluated in this study to examine operational efficiency of banking sector of Bangladesh are profitability and competitiveness within the sector.

*5.3.1. Profitability*

Table 3 presents profitability indicator ROA for sub-sector of banks over the period 1998 to 2011. As it is observed from the Table 3, ROA differ largely by banking sub-sectors even after the reform measures. The ROA of the SCBs were found to be nil during the period 1998 to 2000, which were even worst (negative) in case of the DFIs, -3.7%. The huge loss of the DFI's in 2000 was mainly due to making of provisions by debiting 'loss' in their book of accounts. During 2011, ROA of SCB and DFI have been 1.3% and less than 0.1% respectively.

| Return on Asset (ROA) % | | | | | | | | | | | | | | |
|---|---|---|---|---|---|---|---|---|---|---|---|---|---|---|
| Bank Types | 1998 | 1999 | 2000 | 2001 | 2002 | 2003 | 2004 | 2005 | 2006 | 2007 | 2008 | 2009 | 2010 | 2011 |
| SCBs | 0 | 0 | 0 | 0.06 | 0.10 | 0.08 | -0.14 | -0.10 | 0 | 0 | 0.70 | 1.0 | 1.1 | 1.3 |
| DFIs | -2.8 | -1.6 | -3.7 | 0.67 | 0.33 | -0.04 | -0.13 | -0.13 | -10.15 | -0.27 | -0.60 | 0.4 | 0.2 | 0.1 |
| PCBs | 1.2 | 0.8 | 0.8 | 1.12 | 0.75 | 0.69 | 1.24 | 1.06 | 1.07 | 1.28 | 1.37 | 1.6 | 2.1 | 1.6 |
| FCBs | 4.7 | 3.5 | 2.7 | 2.8 | 2.36 | 2.55 | 3.15 | 3.09 | 3.34 | 3.10 | 2.94 | 3.2 | 2.9 | 3.2 |
| All Banks | 0.3 | 0.2 | 0 | 0.69 | 0.52 | 0.49 | 0.69 | 0.60 | 0.79 | 0.89 | 1.1 | 1.4 | 1.8 | 1.5 |

Table 3: ROA of Banking Sub-sector
Source: Annual Report, Bangladesh Bank

PCB's ROA is found to have a positive but inconsistent trend, whereas the FCB's showed a consistently better trend over the last 14 years. The superior performance of foreign banks might be due to their technological advantage and product differentiation capabilities which might have been eroded to an extent by the local private banks in recent years. The ROA of PCBs and FCBs were strong, 2.1% and 2.9% respectively, in 2010 implying a growing competition between them after the reform measures. Despite benefiting from high interest spread, profitability has been modest for sub-sector of banks due to high operating cost resulting from low level of automation and over-staffing in state-owned banks and inability to increase fee based income and in most recent years mostly due to increased provisioning requirement of Bangladesh Bank.

*5.3.2. Competitiveness within the Sector: Boone Indicator*

Recently, a new approach to measuring competition has been introduced by Boone (Boone, 2008), which is a measure of degree of competition based on profit-efficiency of the banking sector. It is calculated as the elasticity of profits to marginal costs. The rationale behind the indicator is that higher profits are achieved by more-efficient banks.

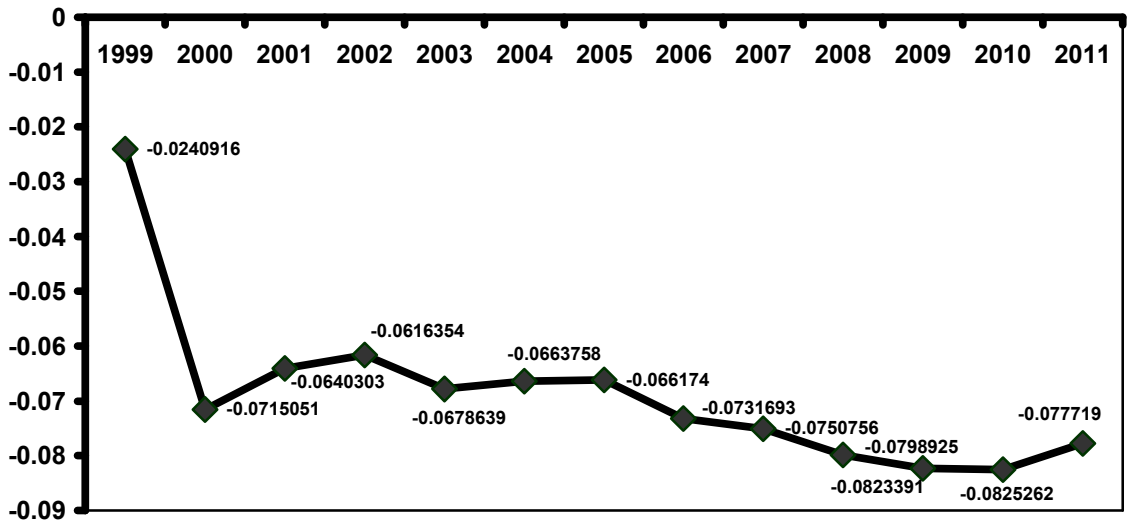

Chart 2: Boone Indicator Index  
Source: Economic Research

      An increase in the Boone indicator implies a deterioration of the competitive behavior of financial intermediaries. Hence, the more negative the Boone indicator, the higher the degree of competition within a sector. In Chart 2, higher negative boon indicator over the period of 1999 to 2011 indicates growing competitive market behavior of the banking industry.

*5.4 Stability*  
Following the framework of World Bank, the stability of banking sector is evaluated in terms of Z score, Asset Quality and Capitalization condition of banking sub-sectors.  
*5.4.1. Z- Score*  
Z-score is a widely used indicator of financial stability in the recent studies. It captures the probability of default of a country's banking system, calculated as a weighted average of the Z-scores of a country's individual banks, where the weights are based on the individual banks' total assets. Z-score compares a bank's buffers, i.e. capitalization and returns, with the volatility of those returns. Intuitively, Z-score can be considered as an inverse measure of insolvency risk, that is, the threat for a bank to be forced out of business because of a lack in capital to compensate for a decline in the value of its' assets (Roy, 1952). A higher Z-score implies a lower probability of insolvency and a greater financial stability.

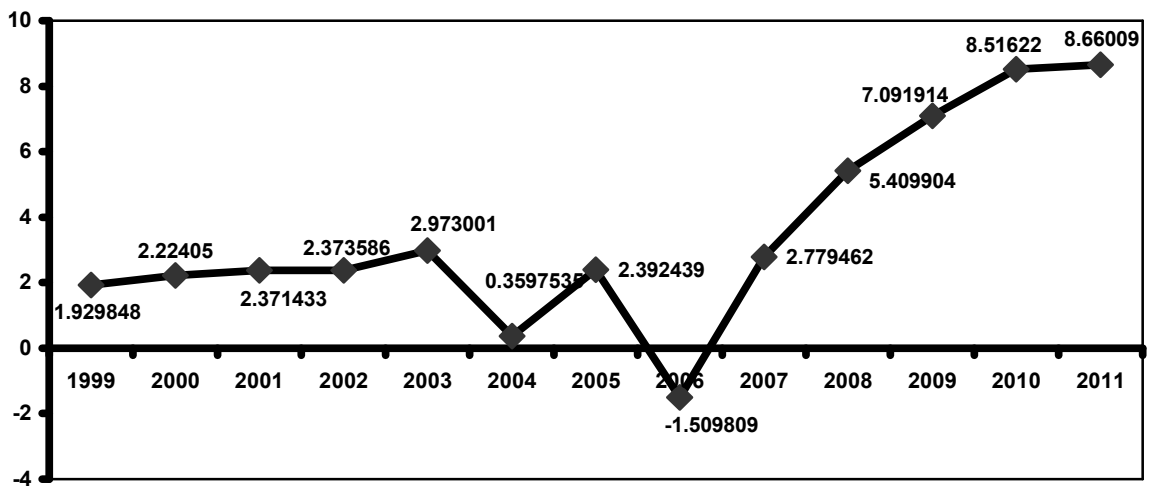

Chart 3: Average Z-Score of all Banks  
Source: Economic Research

      Chart 3 reports the average Z-scores at the bank-year level over the period of 1999 to 2011. The average Z-score across all banks during 2011 is about 8.7, indicating that on an average, profits (ROA) have to fall by 8.7

times their standard deviation to deplete bank's equity. The analysis suggests a modest improvement in financial stability of the banking sector since 2007. Hence, this indicates that banking reforms occurring in terms privatization of state banks and establishment of prudential regulation and supervision led to banks having better opportunity to explore economies of scope and scale and thereby create more stable revenue.

*5.4.2 Asset Quality*

Asset quality indicated by nonperforming loan to total loan ratio has persistently been weak in Bangladeshi banks with nearly one third of the loan portfolio being classified as non-performing loans (NPLs) for the systemically important state-banks and DFIs. At end of 2011, the reported gross NPL to total loan ratios in Table 4 for the SCBs and DFIs were 11.3% and 24.6%, respectively, whereas they were more acceptable at 2.9% for PCBs and FCBs. Meanwhile, the reported banking system gross NPL ratio of 6.1% during 2011 is much improved compared to the highest 41% posted in 1999. The high level of NPLs at SCBs and DFIs is largely attributed to politically directed lending extended on non‐market terms as well as lending under government‐directed schemes. This position is also worsened by the limited credit appraisal, post-disbursement credit monitoring and risk management skills in these institutions. Furthermore, some banks are reluctant to write-off historically bad loans because of the poor quality of underlying collateral and therefore to avoid the recognition of hefty losses on their income statement as well as the legal impediments in recovering loans that are written-off.

| Bank Type | 1997 | 1998 | 1999 | 2000 | 2001 | 2002 | 2003 | 2004 | 2005 | 2006 | 2007 | 2008 | 2009 | 2010 | 2011 |
|---|---|---|---|---|---|---|---|---|---|---|---|---|---|---|---|
| SCBs | 36.6 | 40.4 | 45.6 | 38.6 | 37.0 | 33.7 | 29.0 | 25.3 | 21.4 | 22.9 | 29.9 | 25.4 | 21.4 | 15.7 | 11.3 |
| DFIs | 65.7 | 66.7 | 65.0 | 62.6 | 61.8 | 56.2 | 47.4 | 42.9 | 34.9 | 33.7 | 28.6 | 25.5 | 25.9 | 24.2 | 24.6 |
| PCBs | 31.4 | 32.7 | 27.1 | 22.0 | 17.0 | 16.4 | 12.4 | 8.5 | 5.6 | 5.5 | 5.0 | 4.4 | 3.9 | 3.2 | 2.9 |
| FCBs | 3.6 | 4.1 | 3.8 | 3.4 | 3.3 | 2.6 | 2.7 | 1.5 | 1.3 | 0.8 | 1.4 | 1.9 | 2.3 | 3.0 | 2.9 |
| All Banks | 37.5 | 40.7 | 41.1 | 34.9 | 31.5 | 28.0 | 22.1 | 17.6 | 13.6 | 13.2 | 13.2 | 10.8 | 9.2 | 7.3 | 6.1 |

Table 4: Gross NPL to Total Loan Ratio % (without adjustment for actual provision and interest suspense)
Source: Annual Report, Bangladesh Bank

The decline in the NPL to Total Loan Ratio from 2001 to 2011 was aided by a reduction in absolute NPLs, while in more recent years it was influenced by higher loan growth. Some reduction from the absolute NPL base was also made possible due to write-off of long overdue NPLs, in keeping with guidelines issued by Bangladesh Bank. With nearly 80% of all classified loans in the loss category, the reduction of NPLs through recoveries seems remote hence write-offs appear to be the likely outcome. Asset quality is likely to come under pressure with recoveries becoming tougher, therefore to wipe out unnecessarily and artificially inflated size of balance sheet, uniform guidelines of write-off have been introduced in 2003. According to the policy, banks may, at any time, classify write-off loans as bad or loss. Those loans, which have been classified as bad or loss for the last five years and above and loans for which 100% provisions have been kept, should be written off immediately. Besides loan loss provisioning and write-offs, stronger regulation, enhanced legal powers of the banks to collect problem loans through the Money Loan Court and better screening of new loans facilitated by the Credit Information Bureau have also improved the NPL ratio. The government made progress in this regard by improving the legal framework for debt recovery by enacting and amending Acts from time to time. The government strategy, which is being implemented by the Bangladesh Bank with assistance from the International Monetary Fund (IMF) and the World Bank, includes limiting the annual credit portfolio growth of SCBs to 5%. However, in response to the global financial crisis, the Government in March 2009 doubled the lending limits of the SCBs to 10% to boost domestic investment. Banking sub-sector structural reform is well under way with the assistance of the current Enterprise Growth and Bank Modernization Project and the Poverty Reduction and Growth Facility. These initiatives aim to bring about a competitive private banking system by reforming the SCBs through (Recent Reform Initiatives, 2012).

*5.4.3 Capitalization*

Capital adequacy focuses on the total position of banks' capital and the protection of depositors and other creditors from the potential shocks of losses that a bank might incur. It helps absorbing all possible financial risks like credit risk, market risk, operational risk, residual risk, core risks, credit concentration risk, interest rate risk, liquidity risk, reputation risk, settlement risk, strategic risk, environmental and climate change risk etc. Bangladesh Bank adopted BASEL-I norms in 1996 and it was followed till the end of 2009. Following BASEL-I principles, banks were required to keep Capital Adequacy Ratio (CAR) of not less than 9.00 percent of their risk-weighted assets with at least 4.5 percent in core capital or Taka 1.00 billion whichever is higher. Table 5 shows that aggregate capital adequacy ratio of the banking sector showed a downward trend since 1997 and declined to 5.6% in 2005. However, in 2009, the ratio rose to 11.6%, the highest during the last 13 years. The SCBs could not attain the required level till 2009 and DFIs failed to maintain required CAR except for the year 2004. FCBs have the CAR much above the required standard over the stated period.

| Bank Type | 1997 | 1998 | 1999 | 2000 | 2001 | 2002 | 2003 | 2004 | 2005 | 2006 | 2007 | 2008 | 2009 | 2010 | 2011 |
|---|---|---|---|---|---|---|---|---|---|---|---|---|---|---|---|
| SCBs | 6.6 | 5.2 | 5.3 | 4.4 | 4.2 | 4.1 | 4.3 | 4.1 | -0.4 | 1.1 | 7.9 | 6.9 | 9.0 | 8.9 | 11.7 |
| DFIs | 6.0 | 6.9 | 5.8 | 3.2 | 3.9 | 6.9 | 7.7 | 9.1 | -7.5 | -6.7 | -5.5 | -5.3 | 0.4 | -7.3 | -4.5 |
| PCBs | 8.3 | 9.2 | 11.0 | 10.9 | 9.9 | 9.7 | 10.5 | 10.3 | 9.1 | 9.8 | 10.6 | 11.4 | 12.1 | 10.1 | 11.5 |
| FCBs | 16.7 | 17.1 | 15.8 | 18.4 | 16.8 | 21.4 | 22.9 | 24.2 | 26.0 | 22.7 | 22.7 | 24.0 | 28.1 | 15.6 | 21.0 |
| Total | 7.5 | 7.3 | 7.4 | 6.7 | 6.7 | 7.5 | 8.4 | 8.7 | 5.6 | 6.7 | 9.6 | 10.1 | 11.6 | 9.3 | 11.4 |

Table 5: Capital Adequacy Ratio % (CAR)
Source: Annual Report, Bangladesh Bank

Banking sector of Bangladesh is in a healthy state since capital adequacy related BASEL-II norm has been adopted fully in 2010. Capital adequacy ratio of all banks was 11.4% in 2011 and 9.3% in 2010 as against required CAR level of greater than or equal to 10% as described in BASEL-II accord. Table 5 shows that on 31 December 2011, in aggregate the SCBs, DFIs, PCBs and FCBs maintained CAR of 11.7%, -4.5%, 11.5% and 21.0% respectively. But individually, three PCBs and two DFIs did not maintain the minimum required CAR whereas, all PCBs and FCBs complied with the minimum required capital. Bank's having CAR below the regulatory requirement are categorized as 'problem banks' and are asked to make-up for the shortfall by increasing their paid-up capital. Until recent development in CAR ratio, capitalization levels of banks were poor and affected by weak capitalization in state banks as well as marginal capitalization in some private banks. The weak capitalization levels and inadequate provisioning on a high NPL base resulted in an extremely weak CAR ratio over the stated period. Improvement in CAR ratio that is visible from 2007 was largely enabled by a 'valuation adjustment' made during the conversion of three SCBs (Agrani, Janata and Sonali Bank) into limited liability companies. Clearly, the capital base of the banking sector has tremendously expanded especially during the last four years. This has been possible due to the transfer of a sizeable portion of banks' profits into capital. As a result, the base of banking system has become much stronger. At present, Bangladesh Bank is preparing to adopt and implement the BASEL-III principles in near future. While capitalization could further improve in future, due to the enhanced minimum capital requirements, sustaining such improvements would depend on the ability of the banks to ensure their equity base is not affected by operational losses, especially in a challenging environment with renewed pressure on asset quality (Financial Stability Report, Bangladesh Bank).

**6. Conclusion and Policy Inferences**
In spite of deterioration of some financial soundness indicator banking system continued to show growing resilience over the years. Banking sector penetration enhanced due to bringing unbanked people into banking network and expanding branch network. Banking sector's contribution to economy grew notably, which is evident as share of M3 to GDP reported increment. Banks are becoming more efficient as indicated by ROA, which is repotted to be modest for banking sub-sectors but recorded a decline in 2011 due to creating additional loan loss provision as a result of the new stricter loan loss provision requirement of Bangladesh Bank. Banking sector is becoming more competitive as reported by Boone Indicator. Banking sectors ability to absorb losses and overcoming risk of insolvency is improving over the years as reported by Z-Score. Non-performing loan ratio is showing declining trend due to reported growth in earning asset and also because of writing-off bad debts. At the end of year 2011 most of the banks except DFIs were able to maintain their minimum required CAR of 10% in line with Pillar-I of the Base-II framework. Historically, the performance of banking sector has been weak but this trend has started to reverse gradually with the initiation and enforcement of different reform measures as reported in this study. However, it should be noted that confidence of individual's in banking system as a whole is the key in banking. As banks are vital institutions of financial intermediation in Bangladesh, any loss of trust, may cause a shortage of funds in banking system and inefficient practices would result in a failure of banking system and cause major economic slowdown in Bangladesh. Therefore, this study suggests stricter enforcement of following factors for improvement and stability of banking sector:

- More stringent supervision of banks
- Stringent loan classification and provisioning
- Constraint on loan rescheduling
- Automation of the payment and settlement system
- Strengthening of the Credit Information Bureau (CIB)
- Broad-based social inclusion program
- Emphasis on risk management and internal control framework;
- Satisfactory international reserve
- Implementation of Basel-III framework that focuses on minimum capital requirement, supervisory review process and market discipline.